\documentstyle[12pt]{article}

\begin{document}

\vskip 1.5 cm 
\centerline{\Large 
Space of spaces as a metric space}
\vskip 1.0 cm 
\centerline{\large {\sc Masafumi Seriu}}
\centerline{\it  Department of Physics, Fukui University}
\centerline{\it  Fukui 910-0017, Japan}
\centerline{E-mail:{\it mseriu@edu00.f-edu.fukui-u.ac.jp}}
\centerline{and} 
\centerline{\it Yukawa Institute for Theoretical Physics}
\centerline{\it Kyoto University}
\centerline{\it Kyoto 606-8224, Japan}
\centerline{E-mail:{\it mseriu@yukawa.kyoto-u.ac.jp}}
\vskip 1.5cm

\begin{abstract}
In spacetime physics,  we frequently need to consider 
a set of all  spaces (`universes') as a whole. In particular, 
 the concept of `closeness' between spaces is essential. 
 However there has been no established  mathematical theory so far 
 which deals with a space of spaces in a suitable manner for 
 spacetime physics. 
 
 Based on the scheme of the spectral representation of geometry, 
 we construct a space ${\cal S}_N$, which is 
  a space of all compact Riemannian manifolds equipped with 
  the spectral measure of closeness. We show that ${\cal S}_N$ can be 
  regarded as  a metric space. We also show other desirable
   properties of ${\cal S}_N$, 
   such as the partition of unity, locally-compactness 
  and the second countability. These facts show that 
  the space ${\cal S}_N$ can be a basic arena for spacetime physics. 
\vskip .5cm
Running Title: `{\it Space of spaces as a metric space}'  or 
 `{\it Space of spaces}' 

 \end{abstract}

\section{Introduction}

In the course of the development of theoretical physics, 
mathematics   has  been an efficient language for a precise 
formulation  and analysis of the problems.  
However sometimes physics goes ahead, namely  we are occasionally 
forced  to face with  a problem in physics for 
which appropriate mathematical language has not yet been 
established. In this case, we cannot even state  the problem 
properly, though its importance is obvious.  
Though troublesome,  it is  such  a situation that  can be a strong 
motivation  for a  new development of mathematics, which in turn 
would help  further  progress in theoretical physics. 

The theory presented here can be regarded as  of this kind. 
In spacetime physics, it often happens that  
not a single space (`universe'), but 
a set of spaces  should be considered as a whole. 
However there is no workable mathematical theory 
suitable for handling  such a situation.
   
Perhaps the most famous example of this category
 is  the `spacetime foam' picture due to Wheeler~\cite{WH}: 
At near the Planck scale 
($l_{pl}$=$(\frac{G\hbar}{c^3})^{1/2}$$\simeq$
$10^{-33}$ cm),  it is anticipated that there are    
 drastic topological fluctuations of spacetime taking place 
 (`spacetime foam'),  
because of  the quantum effect on spacetime. 
Now as  the observational energy scale $E$ goes down, 
the finer topological 
structure of scale less than $E^{-1}$ 
would be averaged in some manner, and the effective geometry 
would seem  simpler than the original one. If $E$ is decreased 
to the much lower energy scale, say $10^{15}$ GeV,  almost all of 
the topological handles  would be smoothed out, resulting in 
the  simplest  structure as we experience usually. 

In the 
`spacetime foam' picture briefly described above, 
 we find the concept of a `set of spaces' appears twice: 
 \begin{description}
 \item{(1)}
  Topological fluctuations. We are tacitly considering  
 a set of spaces with various topologies. 
\item{(2)}
 Scale-dependent topology~\cite{VI,SE1,SE2}. 
\end{description}
 It is appropriate to explain here the scale-dependent topology
  in some detail: 
 In the standard mathematical context, 
 topology  by definition is a scale-free concept, i.e. 
 it represents the geometrical properties that are 
 independent from the sizes  of  the  structures. However topology in 
 spacetime physics is quite different:
   Here the observational energy scale $E$ naturally 
 enters  into a discussion, so that geometrical structures 
 of scales less than $E^{-1}$ are  of no significant meaning. 
 Thus  we need to consider effective topology as a function of 
 $E$. In other words, we need to establish a procedure of 
 `topological approximation' at the energy scale $E$.
 In this manner,    not only the topology (in the ordinary sense) 
 of a handle, but also its size  should be taken into account 
 in spacetime physics.    
 Hence  it is desirable  to establish a suitable framework in which 
 local geometry and global topology are treated 
 on the same footing~\cite{SE3}.

As another  example 
which requires the concept of a `set of spaces' 
    we  mention a fundamental problem  in   cosmology. 
 The real geometry  of our  universe is very 
  complicated  
 so that we can perceive how  our universe is  only through  
 cosmological models. 
 Thus cosmology in principle requires a mapping 
 procedure from reality to a model. 
 By analyzing the observational data, we choose  the optimal model 
 among a set of models which is most compatible with the data. 
 Here a mapping from reality to the model, or smoothing out the reality 
 to get the model is taking place, but so far, 
 this process is not understood well. 
 We need a scheme for  analyzing the mapping procedure itself 
 quantitatively and for  judging the validity of  the model choice. 
 For this purpose, we need to establish  a suitable  
 language which describes
  the `closeness' between two spaces (reality and a model). 
 Thus we should  give a definite   meaning to 
 a space of all spaces, supplying  a suitable 
 distance between spaces,  for further understanding of the 
 problem of model fitting in cosmology~\cite{SE4}.

The fundamental problems in spacetime physics like the examples 
mentioned above have been frequently discussed so far,  
without  firm foundations for the space of spaces, effective 
topology, topological approximation,  and so on.

As perhaps the first attempt to describe the scale-dependent topology 
quantitatively, the scattering cross-sections of a small handle of 
various topologies in 2-dimensional space have been 
investigated and it has been analyzed how the cross-sections 
are influenced by  
the  topology of a handle and the energy scale of a probe~\cite{SE1,SE2}.

Based on  this preliminary investigation, the more systematic scheme 
for handling  space structures has been introduced:       
  The spectral representation of geometrical structures 
    along with 
  the  spectral distance between spaces~\cite{SE3}.   
  It echoes a famous question in 
  Riemannian geometry, `Can one hear the shape of a drum?'~\cite{KA}
  The basic idea behind the spectral representation  is very clear: 
  We use  the `sounds' of a space    to characterize 
  the geometrical structures of the space. 
  To be more specific, we use 
  the set of eigenvalues of 
  an elliptic operator (typically, the Laplacian) 
 $\{\lambda_n\}_{n=1}^\infty$  to characterize 
 geometry. (We confine ourselves to the case 
 of spatially compact spaces for definiteness.) 
 Then one can also introduce 
 a measure of closeness $d_N({\cal G},{\cal G}')$ 
 between two geometries  $\cal G$ and ${\cal G}'$ by comparing 
$\{\lambda_n \}_{n=1}^N$ for $\cal G$ and 
$\{\lambda'_n \}_{n=1}^N$ for ${\cal G}'$. (We can treat the 
cut-off number $N$ as a running parameter.) 

There are several advantages for the spectral representation:
\begin{description}
\item{(1)} The spectra $\{\lambda_n\}_{n=1}^\infty$ are 
countable number positive quantities and they are easy to handle. 
\item{(2)} The spectra $\{\lambda_n\}_{n=1}^\infty$ contain 
the information on  
both the local geometry and global topology  of a space. In other words, 
$\{\lambda_n\}_{n=1}^\infty$ are suitable quantities for treating  
the local and global geometry in a unified manner. 
\item{(3)} On dimensional grounds, the lower (higher) spectrum 
corresponds to the larger (smaller) scale behavior of geometry.
For instance,  if we compare two sets of spectra up to the $N$-th 
eigenvalue, 
$\{\lambda_n\}_{n=1}^N$ and $\{\lambda'_n\}_{n=1}^N$, 
we are in effect 
comparing the corresponding two geometries, $\cal G $ and ${\cal G}'$,
neglecting the small-scale behavior of order $o(\lambda_n^{-1/2})$. 
In this manner,  the spectra are suitable for describing the
 scale-dependent behavior of geometry.   
\item{(4)} The spectra $\{\lambda_n\}_{n=1}^\infty$ are 
spatially diffeomorphism invariant quantities, and 
the spectral distance $d_N$  also possesses this property. 
\item{(5)} The spectral distance $d_N$ between spaces is 
constructed from purely `internal' concepts, i.e. 
the spectra are   defined within spaces themselves, and they are 
independent from  the way of embeddings  into some other space.
In this sense $d_N$ is a physical measure of closeness between spaces. 
\end{description}
  
   The basic properties of the spectral distance 
  have been investigated.  Among several possibilities,
   one  particular choice of the spectral distance $d_N$ 
   is especially important since it can be derived from 
   quite a general argument of introducing  a distance, and 
    since it can be related 
    to the reduced density matrix element in quantum cosmology 
   under some circumstances.  At the same time, however,
    it turned out that this form  of $d_N$ does not 
  satisfy the triangle inequality~\cite{SE3}. Though the 
  triangle inequality is not of absolute necessity, its utility 
  clearly makes the arguments efficient and compact, and furthermore 
  it is compatible with our intuitive notion of `closeness'. 
 Thus further investigations on this point have been awaited.

 In this paper we will show that the failure of the triangle inequality 
 of the spectral distance $d_N$ is only a mild one, and in fact 
 we will see that its  slight modification, $\bar{d}_N$, 
 recovers the inequality. With the help of  $\bar{d}_N$, which is a distance 
 in a rigorous sense, we can  investigate the properties of 
 the space ${\cal S}_N$, 
 the  space of all spaces equipped with $d_N$.  Then we will see that, 
 as topological spaces,  
 the space ${\cal S}_N$ is equivalent to the space of all spaces 
 equipped with $\bar{d}_N$, which is a metric space. Thus 
 ${\cal S}_N$ is a metrizable space and it provides us with 
 a notion of `closeness' between spaces in a manner that is 
 compatible with  our intuitive notion of `closeness'.
 In this way it is justified to 
 treat ${\cal S}_N$ as a metric space,  provided that we resort to  
 $\bar{d}_N$ whenever the triangle inequality is needed in the 
 arguments. We will also show that the space ${\cal S}_N$  
  possesses several desirable properties, such as the partition of 
  unity, locally-compactness, the second countability. 
  These observations support us to regard 
  the space ${\cal S}_N$ as  a basic arena for spacetime physics. 
    
  In Section 2, we will introduce the spectral representation and 
  the spectral distance $d_N$. In Section 3, we will consider 
  the space ${\cal S}_N$, the  space of all spaces equipped with 
  $d_N$, and will see that ${\cal S}_N$ can be regarded as 
  a metric space. We will also investigate several other 
  properties of ${\cal S}_N$. Section 4 is devoted to several 
  discussions.

\section{The Spectral Distance}

Let us first recollect the attempts in  Riemannian geometry to define 
 `closeness' between spaces. 
 
There is  the Gromov-Hausdorff distance $d_{GH}(X,Y)$ between 
 two compact metric spaces $X$ and $Y$~\cite{GR,PE}.  
 It is defined by means of isometric embeddings of 
 $X$ and $Y$ into another  metric space $Z$ such  as 
\[
d_{GH}(X,Y):= \inf_{\varphi_1,\varphi_2} d_H (\varphi_1(X), \varphi_2(Y))\ \ .
\] 
Here $d_H$ denotes the Hausdorff distance\footnote{
Let $(Z, d)$ be a metric space.  For $A \subset Z$ and $\epsilon >0$, 
we define the $\epsilon$-neighborhood of $A$ as  
$B(A,\epsilon ):=\{x \in Z | d(x, A) <\epsilon\}$, where 
$d(x,A) := \inf_{y \in A} d(x,y)$. Then the Hausdorff distance between 
two subsets of $Z$,  $A_1$ and $A_2$, is defined as 
$d_H(A_1, A_2)
     := \inf\{ \epsilon | \epsilon >0,  A_1 \subset B(A_2 , \epsilon)
      , A_2 \subset B(A_1 ,\epsilon ) \}$. } 
 on $Z$ 
;  $\varphi_1:X \hookrightarrow Z$ and 
$\varphi_2:Y \hookrightarrow Z$  are  isometric embeddings of  
$X$ and $Y$ respectively, into $Z$. In other words, 
we search for the optimal isometric 
 embeddings such that   
 $X$ and $Y$ overlap `as close as possible' in $Z$, and the distance 
 is defined as the order of  failure in overlapping.      

 Another quantity related to `closeness' between spaces
  is a  norm of Riemannian manifolds due to Petersen~\cite{PE}. Let   
 $(U_\lambda, \phi_\lambda)_{\lambda \in \Lambda}$ be an atlas of 
 a Riemannian manifold $(\Sigma, h)$. 
 If   each patch $U_\lambda$ is chosen to be sufficiently small,    
 the metric tensors w.r.t. (with respect to) the atlas 
 do not change so much within each chart 
 $(U_\lambda, \phi_\lambda)$, 
 namely,  locally it looks like Euclidean to some extent.   
 Now the larger the size of charts becomes 
 the more the metric tensors vary  within each  chart. 
 The norm due to Petersen 
 can be explained as 
  the maximum size of the admissible charts under the condition 
 that  the variation of the metric tensors within  
 each chart lies within  a given range. Its precise definition 
 is very  complicated and we do not go further here. 
 This norm  measures how close a Riemannian manifold 
 is to the Euclidean space,  but it does not provide us 
 with the distance between two manifolds.  
 
 Though these quantities  play a significant role in the 
 convergence theory of Riemannian geometry~\cite{PE}, it seems
  too abstract and complicated  to be directly applied to spacetime physics.

 Thus let us  focus on  another  measure of closeness between spaces 
 which would be  suitable for spacetime physics. 
 We make use of  the eigenvalues  of an elliptic operator 
 on a space 
 (or `{\it spectra}' hereafter). 
 The spectra 
 contain the information of both local geometry 
 and global topology,  and 
 the difference in geometry reflects on   the difference 
 in the spectra.\footnote{
 Geometrical structures are  classified into two categories:
 local geometry and global geometry (global topology),  though they are 
 related to each other and there is no clear separation between them.
 Throughout this paper `geometry' indicates both, i.e. the integrated 
 geometrical properties. We use the symbols $\cal G$, 
 ${\cal G}'$, etc.  to represent geometry as a whole 
 in this broad sense.} 
 Thus, to state symbolically, 
 ``we `hear' the shape of the universe''. Let us call such a 
 representation of geometry in terms of the spectra the 
 {\it spectral representation}, for brevity~\cite{SE3}.

Now let $Riem$ be a space of all $D$-dimensional, 
compact Riemannian manifolds 
without boundaries. 
  Let ${\cal G} =( \Sigma, h)$,  ${\cal G}'=(\Sigma', h')$ $\in Riem$. 
  (We regard them as models of spaces and not spacetimes.) 
 For definiteness we consider only 
 the Laplacian operator $\Delta$ here as an elliptic operator,  
 though the arguments below are quite universal. 
 
  Setting the eigenvalue problem on each manifold 
  \[
  \Delta f =-\lambda f \ \ ,
  \]  
  the set of eigenvalues (numbered in increasing order) is obtained;  
  $\{\lambda_m \}_{m=0}^\infty$ for $\cal G$  and 
  $\{\lambda'_n \}_{n=0}^\infty$ for ${\cal G}'$.  
 
 The first option that one can imagine easily is probably  
 \[
 d_{Euclid}({\cal G}, {\cal G}')
 :=\sqrt{\sum_{n=1}^N (\lambda_n - \lambda'_n )^2} \ \ ,
 \]
 which is similar to the Euclidean distance on ${\bf R}^N$. However 
 from the viewpoint of physics, it is unsatisfactory for two 
 reasons.
 \begin{description}
 \item{ (1)}  The spectra $\{ \lambda_n \}_{n=1}^\infty$ 
 have the physical dimension $[{\rm Length}^{-2}]$, so that  
  $d_{Euclid}$ also has a  dimension $[{\rm Length}^{-2}]$. 
  It introduces a scale into  the theory, which    is not 
   desirable.  For instance, the statement 
  $d_{Euclid} << 1$ becomes meaningless, and rather we should 
  say $d_{Euclid} << l_{pl}^{-2}$, if we choose $l_{pl}$ as 
  a typical length scale. In this way a particular scale 
  (e.g. $l_{pl}$) enters into the discussion. 
  This is unsatisfactory, considering the 
  fundamental nature of the theory of `closeness' between spaces. 
  \item{(2)} Remembering  the arguments of scale-dependent topology 
  (see \S 1), it is clear that, in spacetime physics, 
   the larger scale behavior of geometry 
  is of more importance than the smaller scale behavior. However, 
  looking at the expression of $d_{Euclid}$, the measure 
  $d_{Euclid}$ counts the smaller scale behavior (i.e. 
  $\lambda_n$ with larger $n$) with more importance, which  is 
  unsatisfactory. 
 \end{description}
    
  Thus a simple difference $\lambda_n - \lambda'_n$ is not appropriate 
  for our purpose. Rather we should take the ratio 
  $\frac{\lambda'_n}{\lambda_n}
  =1+\frac{\delta \lambda_n}{\lambda_n}$, which implies that 
  the difference $\delta \lambda_n:= \lambda'_n- \lambda_n$ 
  in the lower spectrum is 
  counted with more importance. 
  
  Hence we  introduce  a measure of closeness between $\cal G$ and 
  ${\cal G}'$ as
\begin{equation}
  d_N ({\cal G},{\cal G}')= \sum_{n=1}^N {\cal F} 
  \left( \frac{\lambda'_n}{\lambda_n} \right)\ \ ,
\label{eq:d_N_general}
\end{equation}
  where ${\cal F}(x)$ ($x>0$) is a suitably chosen function 
  which satisfies   ${\cal F} \geq 0$,    
  ${\cal F}(1/x)={\cal F}(x)$, 
  ${\cal F}(y)>{\cal F}(x)$ if  $y > x \geq 1$. 
  Most of the cases  we also  require   ${\cal F}(1)=0$.  
  (However, see an exceptional case   below (${\cal F}_2$).) 
  Note that the zero  modes $\lambda_0=\lambda'_0=0$ are 
  excluded from the summation in Eq.(\ref{eq:d_N_general}). 
   On dimensional grounds,  $\lambda_n$ with 
  the larger number  $n$ reflects the smaller scale behavior 
  of the geometry. Therefore the cut-off number $N$ indicates 
  the scale up to which  ${\cal G}$ and ${\cal G}'$ are compared. 
  In other words the difference between  the two geometries 
  in the scale  of order $o\left(1/\sqrt \lambda_N \right)$ is neglected. 
   Treating  $N$ as a running parameter,   
   $d_N({\cal G}, {\cal G}')$ as a function of $N$ indicates the 
   coarse grained similarity 
   between  ${\cal G}$ and ${\cal G}'$   at each scale.  
In this way  the spectral measure of closeness 
 gives a natural basis for  analyzing 
 the scale-dependent behavior of geometries. 
 Here we remember that such a  quantitative,  
 scale-dependent description of the  geometrical structures 
 (the scale-dependent topology in particular)
 is very  essential for developing  spacetime physics (see 
 \S 1)~\cite{SE1,SE2}. 
 
   Now there are several possibilities 
  for the choice of $\cal F$ in Eq.(\ref{eq:d_N_general}). 
  Its detailed form  should be   
   determined according to the  features of  geometry 
   that  we are interested in. Among several 
   possibilities, however, there is  
   one especially interesting choice for $\cal F$, which is  
   ${\cal F}_1(x)=\frac{1}{2} \ln \frac{1}{2}(\sqrt{x}+1/\sqrt{x}) $.  
   By this choice, $d_N$ can be related to the reduced density 
   matrix element in quantum cosmology.  
For the details on the derivation and physical interpretation of this choice,  
we refer the reader to  Ref.\cite{SE3}.
Thus  we get~\cite{SE3}
\begin{equation}
d_N ({\cal G},{\cal G}')
=\frac{1}{2} \sum_{n=1}^N \ln \frac{1}{2}
\left(
\sqrt{\frac{\lambda'_n}{\lambda_n}}
+\sqrt{\frac{\lambda_n}{\lambda'_n}}
\right)\ \ .
\label{eq:d_N} 
\end{equation}
This measure of closeness\footnote{
We will see below that it is  justified to regard $d_N$    
as a distance if a suitable care is taken. Until then, however, let us  call it 
a `measure of closeness' for safety. 
} possesses the following properties:
\begin{description}
\item[(I)] $d_N({\cal G}, {\cal G}')\geq 0$,  and 
      $d_N({\cal G}, {\cal G}')=0$ $\Leftrightarrow$ 
       ${\cal G} \sim {\cal G}'$, where $\sim$ means 
       equivalent in the sense of isospectral manifolds\footnote{
 Two non-isometric Riemannian   manifolds  ${\cal G}$ and ${\cal G}'$ 
 are called   {\it isospectral  manifolds} when 
  $\{\lambda_m \}_{m=0}^\infty \equiv  
  \{\lambda'_n \}_{n=0}^\infty$~\cite{MI,KA,CH}. However,  
   the weaker condition 
   $\{\lambda_m \}_{m=0}^N \equiv  
  \{\lambda'_n \}_{n=0}^N$ is enough instead of 
   $\{\lambda_m \}_{m=0}^\infty \equiv  
  \{\lambda'_n \}_{n=0}^\infty$ for the present purpose. 
  }, 
\item[(II)]  $d_N({\cal G}, {\cal G}')=d_N({\cal G}', {\cal G})$ 
\end{description}
  but  it    does not satisfy the triangle 
inequality~\cite{SE3}, 
\begin{description}
\item[(III)]  
$d_N ({\cal G}, {\cal G}')+d_N ({\cal G}', {\cal G}'') \not\geq 
d_N ({\cal G}, {\cal G}'')$.  
\end{description}
However, it turns out that the breakdown of the triangle inequality
is only a mild one in the following sense:
A universal constant $a (>0)$ can be chosen such that 
$d'_N({\cal G},{\cal G}'):=d_N({\cal G},{\cal G}')+a$ 
satisfies the triangle inequality. Here $a$ is independent of 
${\cal G}$ and ${\cal G}'$.

Indeed there is another option for ${\cal F}$ as
${\cal F}_2(x):=\log_2 \left( 
\sqrt{x}+1/\sqrt{x} \right)^{\frac{c}{N}}$. Here $c$ is an arbitrary 
positive constant. 
Note that ${\cal F}_2= \frac{2c}{N\ln2}({\cal F}_1+\frac{1}{2}\ln2)$.
Then  a modified measure of closeness becomes  
\[
\tilde{d}_N ({\cal G},{\cal G}')
= \sum_{n=1}^N \log_2 
\left(
\sqrt{\frac{\lambda'_n}{\lambda_n}}
+\sqrt{\frac{\lambda_n}{\lambda'_n}}
\right)^{\frac{c}{N}}\ \ .
\]
 This measure 
$\tilde{d}_N$ satisfies 
\begin{description}
\item[(I')] $\tilde{d}_N({\cal G}, {\cal G}')\geq c$,  and 
      $\tilde{d}_N({\cal G}, {\cal G}')=c$ $\Leftrightarrow$ 
       ${\cal G} \sim {\cal G}'$, where $\sim$ means 
       equivalent in the sense of isospectral manifolds, 
\item[(II')]  $\tilde{d}_N({\cal G}, {\cal G}')
                =\tilde{d}_N({\cal G}', {\cal G})$,  
\item[(III')]  
$\tilde{d}_N ({\cal G}, {\cal G}')+\tilde{d}_N ({\cal G}', {\cal G}'') 
\geq \tilde{d}_N ({\cal G}, {\cal G}'')$.  
\end{description}
In this case the triangle inequality holds ({\bf (III')}) but the lower bound 
of $\tilde{d}_N$ is $c (> 0)$  and not zero ({\bf (I')}). 
Note that $\tilde{d}_N=\frac{2c}{N\ln2}d_N + c$.
Thus, the precise form of {\bf (III)} turns out to be    
\[
d_N ({\cal G}, {\cal G}')+d_N ({\cal G}', {\cal G}'') + \frac{N}{2}\ln2 \geq 
d_N ({\cal G}, {\cal G}'')\ \ \ .
\]
 Looking at 
{\bf (I)}-{\bf (III)} and {\bf (I')}-{\bf (III')}, 
we see that the measures of closeness introduced 
here are generalizations of an ordinary distance.\footnote{
It is desirable to construct the theory of  a generalized metric space
which is characterized  by  the generalized axioms of distance:
\begin{description}
\item{(a)} $d(p,q)\geq 0$,  and $d(p,q)=0 \Leftrightarrow p=q$.
\item{(b)} $d(p,q)=d(q,p)$.
\item{(c)} $d(p,q) + d(q,r) + c \geq d(p,r)$,  where $c>0$ is a universal 
           constant independent of $p$, $q$ and $r$. 
\end{description}
 }
 
For  later use, we also pay attention to another 
choice for ${\cal F}$ in 
Eq.(\ref{eq:d_N_general}): we can choose  
${\cal F}_0(x):=\frac{1}{2}\ln\max(\sqrt{x},1/\sqrt{x})$, which is 
a slight modification of ${\cal F}_1$. 
Then Eq.(\ref{eq:d_N_general}) becomes 
\begin{equation}
\bar{d}_N({\cal G}, {\cal G}')
=\frac{1}{2}\sum_{n=1}^{N}
\ln\max\left(\sqrt{\frac{\lambda_n'}{\lambda_n}},
           \sqrt{\frac{\lambda_n}{\lambda_n'}} \right) \ \ . 
\label{eq:d_N_bar}
\end{equation}
It is clear that $\bar{d}_N$ satisfies all of the axioms of a distance. 
In particular it satisfies the triangle inequality
\begin{description}
\item[(III'')]
$\bar{d}_N ({\cal G}, {\cal G}')+\bar{d}_N ({\cal G}', {\cal G}'') 
\geq \bar{d}_N ({\cal G}, {\cal G}'')$ \ \ ,  
\end{description}
because of the relation 
$ \max(x,1/x)\max(y,1/y)\geq  \max(xy, 1/xy)$ for 
$x,y >0$. Therefore $\bar{d}_N$ is a distance.

\section{The space of all spaces equipped with the spectral distance}

We introduce  an $r$-ball centered at ${\cal G}$ defined by $d_N$ as
\[
B({\cal G}, r; d_N )
:=\{{\cal G}'\in Riem/_\sim| 
   d_N \left({\cal G}, {\cal G}' \right)<r  \}\ \ .
\]
Here    $d_N$ is the one defined  by Eq.(\ref{eq:d_N}) and 
$\sim$ indicates 
the identification of isospectral manifolds. 

We also consider  
an $r$-ball centered at ${\cal G}$ defined by $\bar{d}_N$ as 
\[
B({\cal G}, r; \bar{d}_N )
:=\{{\cal G}'\in Riem/_\sim| 
   \bar{d}_N \left({\cal G}, {\cal G}' \right)<r  \}\ \ .
\]

Below we will show that the set of all balls defined by $d_N$ 
forms a basis of topology ({\bf Lemma 6} below),  and that 
the  topology generated by this set of balls 
is equivalent to the topology generated by 
the set of all balls defined by $\bar{d}_N$ 
({\bf Theorem 1} below).  Here we note that  the latter 
topology makes the space $Riem/\sim$  a metric space. 
Thus 
  the   space
\begin{equation}    
{\cal S}_N^o
:=\left(Riem, d_N \right)/_\sim\ \ 
\label{eq:G_N}
\end{equation}
turns out to be  a metrizable space, which is an 
idealistic  property. 
 (We will 
consider its completion ${\cal S}_N$ after establishing 
{\bf Theorem 1}.)

Now we prepare a series of {\bf Lemma}'s before showing 
{\bf Theorem 1}.

\begin{description}
\item{\bf Lemma 1}\\ 
\subitem{(1)} $d_N({\cal G}, {\cal G}') \leq \bar{d}_N({\cal G}, {\cal G}')$.
\subitem{(2)} 
{\it
 For $\forall B({\cal G}, r; d_N)$ there  exists 
                $r' (>0)$ s.t. $B({\cal G}, r'; \bar{d}_N)$$\subset$
                $B({\cal G}, r; d_N)$.
} 
\subitem{(3)}    
{\it  
 For $\forall B({\cal G}, r; \bar{d}_N)$ there  exists 
                $r' (>0)$ s.t. $B({\cal G}, r'; d_N)$$\subset$
                $B({\cal G}, r; \bar{d}_N)$.
} 
\subitem{\it Proof:}\\
{\bf (1):} It immediately follows from  the inequality 
$\frac{1}{2}(p+1/p) \leq \max (p, 1/p)$ for $p>0$. 

{\bf (2):}
  Indeed, it follows that 
   $B({\cal G}, r; \bar{d}_N)$$\subset$$B({\cal G}, r; d_N)$  
   due to  {\bf Lemma 1}  (1).

{\bf (3):}   
Suppose there exist ${\cal G}$ and  $r >0$ s.t. 
$B({\cal G}, \epsilon; d_N)$$\not \subset$
$B({\cal G}, r; \bar{d}_N)$ 
for $\forall \epsilon >0$. For a fixed $\epsilon >0$, 
take  ${\cal G}'$ s.t. 
${\cal G}' $ $\in$ $B({\cal G}, \epsilon; d_N) 
\backslash  B({\cal G}, r; \bar{d}_N)$. 
Then   
$d_N ({\cal G},{\cal G}')< \epsilon$, which  implies that  
$\frac{1}{2}\ln \frac{1}{2}
\left(
\sqrt{\frac{\lambda'_n}{\lambda_n}}
+\sqrt{\frac{\lambda_n}{\lambda'_n}}
\right)$$ < \epsilon$ for $n=1,2, \cdots N$.  
(Here $\{\lambda_n \}_{n=0}^\infty$ and 
$\{\lambda'_n \}_{n=0}^\infty$ are the spectra corresponding to 
${\cal G}$ and ${\cal G}'$, respectively.) 
Then it easily follows that 
$1 \leq \max (\sqrt{\frac{\lambda'_n}{\lambda_n}}, 
\sqrt{\frac{\lambda_n}{\lambda'_n}}) < 
\exp 2\epsilon + \sqrt{\exp 4\epsilon -1}$ for $n=1,2, \cdots N$. 
Thus we get  a relation 
$r \leq \bar{d}_N ({\cal G},{\cal G}') <$$ 
\frac{N}{2}\ln (\exp 2\epsilon + \sqrt{\exp 4\epsilon -1})$. 
However,  this inequality  cannot  hold for $\epsilon$ s.t.  
$0 < \epsilon < \frac{1}{2}
\ln(\cosh\frac{2r}{N})$, 
which is a contradiction. 
$\Box$
\end{description}

\begin{description}
\item{\bf Lemma 2}\\ 
{\it
 For $\forall {\cal G}'' \in B({\cal G}', \epsilon; \bar{d}_N )$, 
it follows that 
$0 \leq \frac{|\lambda''_n - \lambda'_n|}{\lambda'_n}$
$< \exp 4\epsilon -1$ $(n=1,2, \cdots, N)$. 
Here $\{\lambda'_n \}_{n=0}^\infty$ and 
$\{\lambda''_n \}_{n=0}^\infty$ are the spectra corresponding to 
${\cal G}'$ and ${\cal G}''$, respectively. 
}

\subitem{\it Proof:}\\
The assumption implies that 
$0 \leq \frac{1}{2}\ln \prod_{n=1}^N 
\max\left(\sqrt{\frac{\lambda''_n}{\lambda'_n}},
           \sqrt{\frac{\lambda'_n}{\lambda''_n}} \right)$$< \epsilon$, or\\  
    $1 \leq \prod_{n=1}^N
     \max\left(\sqrt{\frac{\lambda''_n}{\lambda'_n}},
           \sqrt{\frac{\lambda'_n}{\lambda''_n}} \right)$$< \exp 2\epsilon$. 
Thus $1 \leq \max\left(\sqrt{\frac{\lambda''_n}{\lambda'_n}},
           \sqrt{\frac{\lambda'_n}{\lambda''_n}} \right)$$< \exp 2\epsilon $ 
    ($n=1,2, \cdots, N$).  From this inequality,   
    either  $0 \leq \frac{\lambda''_n - \lambda'_n}{\lambda'_n}$  
   $ < \exp 4\epsilon -1$ or 
   $0 \leq \frac{\lambda'_n - \lambda''_n}{\lambda''_n}$  
   $ < \exp 4\epsilon -1$ follows. Then, it is straightforward to get 
   $0 \leq \frac{|\lambda''_n - \lambda'_n|}{\lambda'_n}$  
   $ < \exp 4\epsilon -1$. 
   $\Box$
\end{description}

\begin{description}
\item{\bf Lemma 3}\\ 
{\it 
Let $\{\lambda_n \}_{n=0}^\infty$, $\{\lambda'_n \}_{n=0}^\infty$
 and $\{\lambda''_n \}_{n=0}^\infty$ are the spectra for 
 $\cal G$, ${\cal G}'$ and ${\cal G}''$, respectively. 

Then
$d_N({\cal G}, {\cal G}'')
= d_N({\cal G}, {\cal G}')$$
+\frac{1}{4}\sum_{n=1}^N\frac{\lambda'_n - \lambda_n}{\lambda_n + \lambda'_n}$
$\frac{\lambda''_n - \lambda'_n}{\lambda'_n}$
$ + R$.
Here 
$R= \sum_{n=1}^N  c_n 
(\frac{\lambda''_n - \lambda'_n}{\lambda'_n})^2$, with  
$c_n$'s being  finite constants. Furthermore $R$ is bounded  as 
$|R| < \sum_{n=1}^N  
(\frac{\lambda''_n - \lambda'_n}{\lambda'_n})^2 $.   
}

\subitem{\it Proof:}\\
$(1^\circ)$
  Let $f(x):= \frac{1}{2}\ln \frac{1}{2} (\sqrt{x}+1/\sqrt{x})$ ($x>0$). 
 By Taylor-Maclaurin theorem, it follows that  
$f(x+\delta x) = f(x) 
+ \frac{1}{4} \frac{\sqrt{x}- 1/\sqrt{x}}{\sqrt{x}+ 1/\sqrt{x}}
\frac{\delta x}{x}$$+ c(x+\xi \delta x) (\frac{\delta x}{x})^2$ with     
$0 < \exists \xi < 1$.  Here  
$c(x)$  is a  smooth  function  for $x>0$.    

Indeed,  
setting $\sqrt{x}=\exp \theta$, $f(x)$ can be represented as 
$f(x)=\frac{1}{2}\ln \cosh \theta$. Then 
$\frac{df(x)}{dx}=\frac{1}{4x} \tanh (\theta)$$=:\frac{1}{x}F_1 (\theta)$, 
and 
$\frac{d^{(2)}f(x)}{dx^2}=\frac{1}{x^2}F_{2}(\theta) $, 
where $F_2(\theta)=\frac{1}{2}F_1'(\theta)-  F_1(\theta)$
$=\frac{1}{8}(\frac{1}{\cosh^2 \theta}- 2 \tanh \theta)$. 
Thus $\tilde{F}_2(x)$, which is 
$F_2$  regarded as a function of $x$,  
  is  a well-defined function of 
$\sqrt{x}+ 1/\sqrt{x}$ and $\sqrt{x}- 1/\sqrt{x}$ and it is smooth for 
$x>0$. Then we can set  $c(x)=\frac{1}{2!}\tilde{F}_2(x)$. 
We also note that  $|c(x)|=\frac{1}{2}|F_2(\theta)|$
$<\frac{1}{16}(\frac{1}{\cosh^2 \theta} + 2 |\tanh \theta|)$
$< \frac{3}{16}<1$.   

$(2^\circ)$  Applying the result of $(1^\circ)$ to 
 $d_N ({\cal G}, {\cal G}'')$
$=\sum_{n=1}^N f(\frac{\lambda''_n}{\lambda_n})
=\sum_{n=1}^N 
f(\frac{\lambda'_n}{\lambda_n} 
+ \frac{\lambda''_n - \lambda'_n}{\lambda_n}) $,  the claim follows. 
$\Box$
\end{description}

\begin{description}
\item{\bf Lemma 4}\\ 
{\it 
For $\forall {\cal G}' \in B({\cal G}, r; d_N )$,  there exists 
$\epsilon >0$ s.t. 
$B({\cal G}', \epsilon ; \bar{d}_N ) \subset B({\cal G}, r; d_N )$. 
}
\subitem{\it Proof:}\\
Let $\rho := d_N({\cal G}, {\cal G}')$.  (Then $0 \leq \rho < r$.)
For a fixed $\epsilon$, 
take $\forall {\cal G}'' \in B({\cal G}', \epsilon ; \bar{d}_N )$. 
Then,
\begin{eqnarray*}
d_N ({\cal G}, {\cal G}'')
&=& d_N({\cal G}, {\cal G}') 
   +\frac{1}{4}\sum_{n=1}^N
        \frac{\lambda'_n - \lambda_n}{\lambda_n + \lambda'_n}
         \frac{\lambda''_n - \lambda'_n}{\lambda'_n}
       + R \ \  ({\bf Lemma\  3}) \\
&<& \rho + \frac{1}{4}\sum_{n=1}^N 
   \frac{| \lambda''_n - \lambda'_n |}{\lambda'_n} + |R| \ \ .
\end{eqnarray*}   
Now let us pay attention to the last line. 
The last term is bounded as 
$|R| < \sum_{n=1}^N  
(\frac{\lambda''_n - \lambda'_n}{\lambda'_n})^2 $ ({\bf Lemma 3}).   
Due to {\bf Lemma 2}, thus,   
one can choose $\epsilon$  sufficiently small 
  s.t. the last term $|R|$  
is less in magnitude than the middle term. Let 
$\tilde{\epsilon} (>0)$ be such $\epsilon$. Then we can continue 
the  estimation of  $d_N ({\cal G}, {\cal G}'')$ as  
\begin{eqnarray*}
d_N ({\cal G}, {\cal G}'')
&<& \rho + \frac{1}{2}\sum_{n=1}^N 
   \frac{| \lambda''_n - \lambda'_n |}{\lambda'_n}\\
&<& \rho + \frac{N}{2}(\exp 4\tilde{\epsilon} -1)\ \  ({\bf Lemma\  2})\ \ .    
\end{eqnarray*} 
By  choosing  $\tilde{\epsilon}$ again if necessary, 
 we can assume that     
$0 < \tilde{\epsilon} < 
\frac{1}{4}\ln (1+ \frac{2}{N}(r-\rho))$.   
Then it follows that $d_N ({\cal G}, {\cal G}'')< r$ for 
$\forall {\cal G}'' \in B({\cal G}', \tilde{\epsilon} ; \bar{d}_N )$.
 Hence there exists $\tilde{\epsilon} (>0)$ s.t. 
$B({\cal G}', \tilde{\epsilon} ; \bar{d}_N ) \subset B({\cal G}, r; d_N )$. 
$\Box$
\end{description}

\begin{description}
\item{\bf Lemma 5}\\ 
{\it 
For $\forall {\cal G}' \in B({\cal G}, r; \bar{d}_N )$,  there exists 
$\epsilon >0$ s.t. 
$B({\cal G}', \epsilon ; d_N )$$\subset$$B({\cal G}, r; \bar{d}_N )$. 
}
\subitem{\it Proof:}\\
Since $\bar{d}_N$ is a distance, there exists $\epsilon_0 (>0)$ s.t. 
$B({\cal G}', \epsilon_0 ; \bar{d}_N )$
$\subset B({\cal G}, r; \bar{d}_N )$.  However, 
according to {\bf Lemma 1} (3),  
there exists $\epsilon_1 (>0)$ s.t. 
$B({\cal G}', \epsilon_1 ; d_N ) $$\subset$
$B({\cal G}', \epsilon_0; \bar{d}_N)$. Hence 
$B({\cal G}', \epsilon_1 ; d_N ) $$\subset$
$B({\cal G}, r; \bar{d}_N)$. 
$\Box$
\end{description}

\begin{description}
\item{\bf Lemma 6}\\
{\it 
The set of balls 
$\{ B({\cal G}, r; d_N)|\ {\cal G}\in Riem/_\sim ,\ r>0 \}$
can  form a basis of topology. 
}
\subitem{\it Proof:}\\
 Let  $B_1$ and $B_2$ are two balls defined by $d_N$,  and suppose 
 $B_1 \bigcap B_2 \neq \emptyset$. Because of {\bf Lemma 4}, 
  for $\forall {\cal G} \in B_1 \bigcap B_2$,  there exists 
 $r(>0)$ s.t. $B({\cal G}, r; \bar{d}_N) \subset B_1 \bigcap B_2$. 
 However, due to {\bf Lemma 1} (3), there exists $r' (>0)$ s.t. 
$B({\cal G}, r'; d_N)$$\subset$$B({\cal G}, r; \bar{d}_N)$. 
Hence $B({\cal G}, r'; d_N)$$\subset$$B_1 \bigcap B_2$. 
Thus the  set of all balls defined by $d_N$  satisfies 
the condition for a basis of topology.   
$\Box$
\end{description}

Now  we can show  
\begin{description}
\item{\bf{Theorem 1}} \\
{\it
 The set of balls 
$\{ B({\cal G}, r; d_N)|\ {\cal G}\in Riem/_\sim ,\ r>0 \}$
and the set of balls 
$\{ B({\cal G}, r; \bar{d}_N)|\ {\cal G}\in Riem/_\sim,
 \ r>0 \}$ generate  the same topology on $Riem/_\sim$.
}
\subitem{\it Proof:}\\ 
$\forall B({\cal G}, r; d_N)$ is open in $\bar{d}_N$ topology 
due to {\bf Lemma 4}. On the other hand, 
$\forall B({\cal G}, r; \bar{d}_N)$ is open in $d_N$ topology 
due to {\bf Lemma 5}.   
$\Box$
\item{\bf Corollary} \\
{\it 
The space ${\cal S}_N^o$ is a metrizable space. The distance 
function for metrization is provided by $\bar{d}_N$.
}
\end{description}

Hence it is appropriate  to extend ${\cal S}_N^o$ to its completion 
${\cal S}_N$. We understand 
$\{\lambda_n\}_{n=0}^\infty$,  $d_N$ and $\bar{d}_N$ are extended 
on ${\cal S}_N$ accordingly.

Now the metrizable space ${\cal S}_N$ is a normal space, not to mention 
a Hausdorff space\footnote{
 For the basics of point set topology, see e.g. Ref.~\cite{KO,KE,YA}}.  

Due to {\bf Theorem 1}, it is justified to regard 
${\cal S}_N$ as a metric space,  provided that 
we resort to the distance function $\bar{d}_N$ whenever 
the triangle inequality  is needed in a discussion. 
  From now on we call $d_N$ in Eq.(\ref{eq:d_N_general})  
  (the form of $d_N$ in Eq.(\ref{eq:d_N}) in particular) 
 a {\it spectral distance} for brevity.

 Here it may be appropriate to add some comments on  
 the applications of ${\cal S}_N$ to physics. 
From the viewpoint of the practical applications in spacetime physics, 
$d_N$ is more convenient than $\bar{d}_N$ 
since the former is easier to handle than the latter, which contains 
$max$ in the expression. Furthermore  
 $d_N$ can be related to 
the reduced density matrix element for the universe in the context of 
quantum cosmology~\cite{SE3}: It is possible to state that, under
 some circumstances,  two universes $\cal G$ and ${\cal G}'$ 
are separated far in $d_N$ when their quantum decoherence is strong. 

With these comments in mind let us now  turn back 
to the mathematical aspects of the space ${\cal S}_N$.
 
Since ${\cal S}_N$  is a metrizable space, 
it follows that~\cite{KE,YA} 
\begin{description}
\item{\bf Theorem 2} \\
{\it
  The space  ${\cal S}_N$ is paracompact.
}
\end{description}
Since ${\cal S}_N$ is Hausdorff and paracompact, 
it follows that~\cite{KE,YA} 
\begin{description}
\item{\bf Corollary} \\
{\it 
There exists a partition of unity subject to 
 any open covering of ${\cal S}_N$.
}
\end{description} 

Now we prepare two {\bf Lemma}'s to show that 
 ${\cal S}_N$ is locally compact ({\bf Theorem 3} below).

\begin{description} 
\item{\bf Lemma 7}\\
{\it 
 The set 
 $D({\cal G},r; d_N)
   := \{ {\cal G}' \in {\cal S}_N | d_N ({\cal G}, {\cal G}') \leq r \}$
  is closed and compact in ${\cal S}_N$. 
}
  \subitem{\it Proof:}\\
  $(1^\circ)$
Note that 
the map  $d_N ({\cal G}, \cdot )$$: {\cal S}_N \rightarrow [0,\infty)$ 
is continuous  and that $[0,r]$ is closed in  $[0,\infty)$.
Since $D({\cal G},r; d_N)$ is the inverse image of the closed set $[0,r]$
by the continuous map $d_N ({\cal G}, \cdot )$, it is closed in ${\cal S}_N$.

  $(2^\circ)$ We now show that $D({\cal G},r; d_N)$ is 
  sequentially compact. 
  
  Any sequence $\{ {\cal G}_n \}_{n=1}^\infty$ $\subset$ $D({\cal G},r; d_N)$
   can be  embedded into an  $N$-cube in ${\bf R}^N$, 
   $\{ {\cal G}_n \}_{n=1}^\infty \hookrightarrow $$[0,L]^N$ for 
   some $L>0$.  Indeed let 
  $\{\lambda^{(n)}_k \}_{k=1}^\infty$ be 
  the  spectra (zero-mode is excluded) 
  for   ${\cal G}_n$.  
   Then a map 
   $\{\lambda^{(n)}_k \}_{k=1}^N$$\mapsto$
   $\vec{\mu}^{(n)}
   :=\left(\sqrt{\frac{\lambda^{(n)}_1}{\lambda_1}}, 
   \sqrt{\frac{\lambda^{(n)}_2}{\lambda_2}}, \cdots, 
    \sqrt{\frac{\lambda^{(n)}_k}{\lambda_k}}, \cdots, 
     \sqrt{\frac{\lambda^{(n)}_N}{\lambda_N}} \right)$
     provides the  embedding. 
     ($\{\lambda_k \}_{k=1}^\infty$ is the spectra for $\cal G$.) 
     Here we note that 
     $\sqrt{\lambda^{(n)}_k}$ $(k=1,2, \cdots, N)$ is  bounded 
     as $\sqrt{\lambda^{(n)}_k} \leq 
     (\exp 2r + \sqrt{\exp 4r -1}) \sqrt{\lambda_k}$   
    because $d_N({\cal G}, {\cal G}_n) \leq r$. 
    Hence we can set $L=\exp 2r + \sqrt{\exp 4r -1}$. 
     
     Now $\{ \vec{\mu}^{(n)}\}_{n=1}^\infty$ is a sequence in 
     a compact set $[0,L]^N$, so that there exists 
     its subsequence $\{ \vec{\mu}^{(n')}\}_{n'=1}^\infty$ which 
     converges to a point in $[0,L]^N$ (in the sense of ${\bf R}^N$ topology).
    Let this convergent point be 
    $\vec{\mu}^{(\infty)} $
    $=(\mu^{(\infty)}_1, \mu^{(\infty)}_2, \cdots, \mu^{(\infty)}_N )$. 
     
   Then for any $\epsilon$ 
   ($0 < \epsilon < 1$)
    there exists $M \in$$\bf N$ s.t. for 
   $\forall m \geq M$, it follows that 
    $1-\epsilon < \frac{\mu^{(m)}}{\mu^{(\infty)}} < 
    1 + \epsilon$.
   Then it follows that 
  $ \frac{1}{2}\sum_{k=1}^N 
   \ln \max (\frac{\mu^{(m)}}{\mu^{(\infty)}},
                 \frac{\mu^{(\infty)}}{\mu^{(m)}})$
$ < \frac{N}{2}\ln(\frac{1}{1-\epsilon})$. 
   This  implies that, 
   for any $\epsilon$ 
   ($0 < \epsilon < 1$), 
    there exists $M \in$$\bf N$ s.t. for 
   $\forall m, m' \geq M$, 
   $\bar{d}_N({\cal G}_m, {\cal G}_{m'})$
   $ < N \ln(\frac{1}{1-\epsilon})$. Namely 
   $\{ {\cal G}_{n'} \}_{n'=1}^\infty$ 
   corresponding to $\{ \vec{\mu}^{(n')}\}_{n'=1}^\infty$ 
   is a Cauchy sequence w.r.t. $\bar{d}_N$. 
   Thus $\{ {\cal G}_{n'} \}_{n'=1}^\infty$  is a Cauchy sequence 
   w.r.t. $d_N$ also, due to {\bf Lemma 1} (1).  
   However $D({\cal G}, r; d_N)$ is closed in the complete space 
   ${\cal S}_N$, so that
   $\{ {\cal G}_{n'} \}_{n'=1}^\infty$ converges to a point 
    $\exists {\cal G}_\infty$ in  $D({\cal G}, r; d_N)$. 
   Hence   any sequence 
   $\{ {\cal G}_n \}_{n=1}^\infty$ $\subset$ $D({\cal G},r; d_N)$ 
   contains a subsequence which converges to a point 
   in $D({\cal G},r; d_N)$, i.e.   $D({\cal G},r; d_N)$ 
   is sequentially compact.
   
   $(3^\circ)$ Since $D({\cal G},r; d_N)$ is a set 
   in the metrizable space ${\cal S}_N$ as well as  
   sequentially compact, it is compact.
$\Box$   
\end{description}

\begin{description} 
\item{\bf Lemma 8}\\
{\it
  The set 
 $D({\cal G},r; \bar{d}_N)
 := \{ {\cal G}' \in {\cal S}_N | \bar{d}_N ({\cal G}, {\cal G}') \leq r \}$
  is closed and compact in ${\cal S}_N$.
}
  \subitem{\it Proof:}\\
Since $\bar{d}_N$ is a distance, it is clear that 
$D({\cal G},r; \bar{d}_N)$ is closed in $\bar{d}_N$-topology. Thus 
$D({\cal G},r; \bar{d}_N)$ is closed in ${\cal S}_N$ 
due to {\bf Theorem 1}.   

The rest goes almost in the same manner as the {\it Proof} of {\bf Lemma 7}:   
Any sequence $\{ {\cal G}_n \}_{n=1}^\infty$ $\subset$ 
$D({\cal G},r; \bar{d}_N)$
   can be  embedded into an  $N$-cube in ${\bf R}^N$, 
   $\{ {\cal G}_n \}_{n=1}^\infty \hookrightarrow $$[0,L]^N$ for 
   some $L>0$. The embedding map
   $\{\lambda^{(n)}_k \}_{k=1}^N$$\mapsto$
   $\vec{\mu}^{(n)}$ is the same as in the {\it Proof} of {\bf Lemma 7}. 
   Since $\bar{d}_N({\cal G}, {\cal G}_n) \leq r$, 
   it follows that   $\sqrt{\lambda^{(n)}_k} \leq  \exp 2r\   \sqrt{\lambda_k}$ 
   $(k=1,2, \cdots, N)$ due to {\bf Lemma 2}.  Thus  we can set $L=\exp 2r$.  
    Because  $\{ \vec{\mu}^{(n)}\}_{n=1}^\infty$ is a sequence in 
    a compact set $[0,L]^N$, there exists 
   its subsequence $\{ \vec{\mu}^{(n')}\}_{n'=1}^\infty$ which 
   converges to a point $\vec{\mu}^{(\infty)} $ 
   in $[0,L]^N$ (in the sense of ${\bf R}^N$ topology). Repeating the 
  same argument as in the {\it Proof} of {\bf Lemma 7}, we conclude that 
   $\{ {\cal G}_{n'} \}_{n'=1}^\infty$ 
   corresponding to $\{ \vec{\mu}^{(n')}\}_{n'=1}^\infty$ 
   is a Cauchy sequence w.r.t. $d_N$. 
   However $D({\cal G}, r; \bar{d}_N)$ is closed in the complete space 
   ${\cal S}_N$, then 
   $\{ {\cal G}_{n'} \}_{n'=1}^\infty$ converges to a point 
    $\exists {\cal G}_\infty$ in  $D({\cal G}, r; \bar{d}_N)$. 
   Hence   $D({\cal G},r; \bar{d}_N)$ 
   is sequentially compact.
    Since $D({\cal G},r; d_N)$ is a set 
   in the metrizable space ${\cal S}_N$ as well as  
   sequentially compact, it is compact.
$\Box$   
\end{description}

\begin{description}
\item{\bf Theorem 3}\\
{\it 
The space ${\cal S}_N$ is locally compact.
}

Due to  this property of ${\cal S}_N$,  an integral can be constructed  on 
${\cal S}_N$~\cite{LAN}, which is important  to introduce, e.g.,  
 probability distributions over ${\cal S}_N$.

\subitem{\it Proof:}\\ 
For any ${\cal G} \in {\cal S}_N$,  one can take $D({\cal G},r; d_N)$ 
 or $D({\cal G},r; \bar{d}_N)$ as its compact neighborhood 
 because of {\bf Lemma 7} and {\bf Lemma 8}. 
$\Box$
\item{\bf Corollary}\\ 
{\it 
If a sequence of continuous functions on ${\cal S}_N$, 
$\{ f_n \}_{n=1}^\infty$, 
pointwise converges  to a function $f_\infty$, then 
$f_\infty$ is continuos on a dense subset of ${\cal S}_N$. 
}
\subitem{\it Proof:}\\
The set of discontinuous points of $f_\infty$ is a set of 
first category\footnote{A  set of first category is 
  defined as  a set which can be expressed as 
   a union of at most countable number of sets that are 
  nowhere dense.}. 
Since  ${\cal S}_N$ is Hausdorff and locally compact, 
it becomes a Baire space, i.e. a space in which the complement of 
  any  set of first category 
  becomes dense~\cite{KE,YA}. Thus the claim follows.
$\Box$
\end{description}

We also note that, 
 since ${\cal S}_N$ is Hausdorff and locally compact, one can consider 
its one-point compactification ${\cal S}_N \bigcup \{\infty\}$, 
which is Hausdorff. 
Moreover, ${\cal S}_N$ is metrizable so that it is completely regular, then 
 one can construct its  Stone-{\v C}ech compactification~\cite{KE,YA}.

Furthermore we can show 
\begin{description}
\item{\bf Theorem 4} \\
{\it 
The space ${\cal S}_N$ satisfies the second countability axiom.
}
\subitem{\it  Proof:}\\
$(1^\circ)$ 
Since ${\cal S}_N$ is a metrizable space,  
it suffices to show that ${\cal S}_N$ is separable~\cite{KE,YA}. 
First we choose a suitable countable subset of ${\cal S}_N$.  

For a fixed $M\in {\bf N}$, the label 
$(m_1 \leq m_2 \leq \cdots \leq m_N)$ is  
uniquely assigned to the spectra $\{ \lambda_n\}_{n=1}^\infty$, 
where $m_1, m_2, \cdots, m_N \in {\bf N}$: This can be 
 achieved  by choosing $m_1, m_2, \cdots, m_N$ s.t. 
$\frac{m_{n}-1}{M} < \lambda_n {\ell}^2 
\leq \frac{m_{n}}{M}$ ($n=1,2, \cdots N$). 
(Here $\ell$ is any constant  of physical dimension 
$[{\rm Length}]$. 
It has been introduced only for the physical comfort, and it is 
not essential for the arguments below.)  
For a given $M$, thus,    
the space ${\cal S}_N$  is uniquely decomposed into classes labeled 
by  $(M; m_1 \leq m_2 \leq \cdots \leq m_N)$. 
(Some of the classes can be empty.) 
 Then we can  
  choose a representative 
 ${\cal G}_{(M; m_1 \leq m_2 \leq \cdots \leq m_N)}$ in 
 the class $(M; m_1 \leq m_2 \leq \cdots \leq m_N)$
 $\neq \emptyset$. 
 Thus we obtain a countable subset 
 ${\cal C}:=\{ {\cal G}_{(M; m_1 \leq m_2 \leq \cdots \leq m_N)}|
  M, m_1, m_2, \cdots, m_N \in {\bf N} \}$
 (For notational simplicity let  $(M;\vec{m})$ denote  
 $(M; m_1 \leq m_2 \leq \cdots \leq m_N)$ hereafter.)   
 
 $(2^\circ)$ 
Take  $\forall {\cal G}$$\in {\cal S}_N$. For $\forall M \in {\bf N}$, 
there uniquely exists a class $(M; \vec{m})$ s.t. 
${\cal G} \in$$(M; \vec{m})$. Let  $\{\lambda_n \}_{n=0}^\infty$ 
  and $\{\lambda^*_n \}_{n=0}^\infty$ are, respectively, 
   the spectra for 
  $\cal G$ and ${\cal G}_{(M; \vec{m})}$, the representative of the class
  $(M; \vec{m})$.  
 Then  $|\lambda^*_n - \lambda_n|{\ell}^2 \leq \frac{1}{M}$. 
 Hence $\sqrt{1- \frac{1}{M \lambda_n {\ell}^2}} 
    \leq \sqrt{\frac{\lambda^*_n}{\lambda_n}} 
    \leq \sqrt{1+ \frac{1}{M \lambda_n {\ell}^2}} $. 
    Thus 
 $\bar{d}_N ({\cal G}_{(M; \vec{m})}, {\cal G})$
$< \frac{N}{2}\ln \left(1/\sqrt{1-\frac{1}{M \lambda_1 {\ell}^2}}\right)$. 
  Due to {\bf Lemma 1} (1), thus, 
  it follows that $\forall \epsilon >0$, there exists  
  ${\cal G}_{(M; \vec{m})} \in {\cal C}$ s.t. 
  ${\cal G}_{(M; \vec{m})}$ $\in  B({\cal G}, \epsilon;  d_N)$. 
 Hence ${\cal S}_N$ is separable, so that  the claim follows immediately.   
  $\Box$
\end{description}

  {\bf Theorem 1}-{\bf Theorem 4} and 
   their  {\bf Corollary}'s  
   indicate  that one can construct calculus theory on 
  ${\cal S}_N$ to a great extent, which makes ${\cal S}_N$ 
  a basic arena for spacetime physics. Thus it is essential 
  to investigate the mathematical structures of
  ${\cal S}_N$ 
  in detail.

\section{Discussion}

We have  introduced the space ${\cal S}_N$ and 
have shown that  it has  several desirable  properties. In particular 
we have shown that  ${\cal S}_N$
 is a metrizable space and in effect it can be regarded as 
a metric space provided that 
 care is taken with regard to  the triangle inequality: Whenever we 
need the arguments linked with the triangle inequality, it is 
safer to resort to $\bar{d}_N$, which is a slight modification of 
$d_N$ and which defines the same point set topology as $d_N$ 
({\bf Theorem 1}). 
However in most of the cases, $d_N$ is of more importance as well as 
easier to handle in practical applications.  Therefore it is 
significant that it has been justified to treat ${\cal S}_N$ as 
a metric space. 

Several properties of ${\cal S}_N$ that we have shown indicate that 
the space of spaces ${\cal S}_N$ 
provides us with a firm platform for pursuing meaningful 
investigations in spacetime physics (Recall the arguments in \S 1). 
Hence it is awaited that  more detailed investigation on the 
properties of ${\cal S}_N$ would be performed.

Since the spectral distance is explicitly defined in terms of the 
spectra, that are of  definite physical meaning, it possesses  
direct applicability to physics  as well as  theoretical firmness.  
Explicit applications of the spectral formalism would be discussed 
elsewhere. (See e.g. Ref.\cite{SE4}.)

Finally we make some comments on 
 the isospectral manifolds~\cite{MI,KA,CH}. It  is no surprise 
 that there exist Riemannian manifolds with identical spectra of 
 the Laplacian even though they are non-isometric to each other: 
 In this  case we are comparing `sounds' produced by a single 
 type of oscillation corresponding to the Laplacian. If we change 
 a type oscillation, namely if we use a different elliptic operator, 
 the difference in sound would make a distinction between such spaces.
 If by any chance  there were non-isometric Riemannian manifolds, s.t.   
 the  spectra are identical for any elliptic operator, they should have been
  regarded as identical from the physical point of view.  
   
   We can even imagine a new  picture of `space'  suggested  by 
   the spectral formalism, in which  
   one  regards the whole of 
    the  geometrical information of a space as a collection of 
   all spectral information such as 
\[
Space = 
\bigcup_k \large({\cal D}_k, \{\lambda^{(k)}_n\}_{n=0}^\infty,
               \{ f^{(k)}_n \}_{n=0}^\infty    \large)\ \ , 
\]
  where ${\cal D}_k$ denotes an elliptic operator,   
  $\{\lambda^{(k)}_n \}_{n=0}^\infty$ and 
  $\{f^{(k)}_n\}_{n=0}^\infty$ are  its spectra and eigenfunctions. 
  Here the index $k$ 
  runs over all possible elliptic operators. 
  Any observation selects out   a subclass of elliptic operators related to 
  the observational apparatus 
   so that only a small portion of the whole geometrical 
  information is obtained by a single   observation. 
   In some cases, such incomplete information is not 
    enough to distinguish some class of manifolds. 
    (This is the physical interpretation of isospectral manifolds.) 
     Then one should  
   perform  other type of observations (corresponding to 
   other elliptic operators)   to make  finer distinctions.
   It is also tempting to regard  the  spectral information 
   as most   fundamental.  
   Further investigations are required as to 
   whether this  viewpoint of spacetime makes sense.

\vskip 1cm

The author thanks S. Naito for valuable discussions on point set topology.
He is also grateful to  H. Kodama, K. Piotrkowska and S. Yasukura for helpful 
comments. This work has been completed during the author's stay  
in  Department of Mathematics and Applied Mathematics, 
 University of Cape Town. The author thanks  the department 
for its hospitality.   
He also thanks 
Inamori Foundation, Japan, for encouragement 
as well as financial support. 

\end{document}